\documentstyle[12pt,fleqn,epsf]{article}
\def\beq{\begin{equation}}
\def\eeq{\end{equation}}
\def\Ps{{\mit\Psi}}
\def\pb{\mbox{\raisebox{-.8ex}{\footnotesize{\sc pb}}}}
\def\half{\mbox{$1\over2$}}
\def\2{\mbox{$3\over2$}}
\def\4{\mbox{$3\over4$}}
\def\8{\mbox{$3\over8$}}
\def\3{\mbox{$2\over3$}}
\def\7{\mbox{$8\over3$}}
\def\9{\mbox{$4\over3$}}

\begin{document}
\newcommand{\av}[1]{\langle #1\rangle}
\renewcommand{\thefootnote}{\fnsymbol{footnote}}

\begin{center}
{\Large{\bf Critique of the Wheeler-DeWitt equation}\footnote{To appear
in: {\it On Einstein's Path\/}, ed. by A. Harvey (Springer, 1998).}}
\\[10mm]
Asher Peres\\[7mm]
{\sl Department of Physics, Technion---Israel Institute of
Technology, 32 000 Haifa, Israel}\\[10mm]

{\bf Abstract}\smallskip\end{center}

\begin{quote}
The Wheeler-DeWitt equation is based on the use of canonical
quantization rules that may be inconsistent for constrained dynamical
systems, such as mini\-super\-spaces subject to Einstein's equations.
The resulting quantum dynamics has no classical limit and it suffers
from the infamous ``problem of time.'' In this article, it is shown how
a dynamical time (an internal ``clock'') can be constructed by means of
a Hamilton-Jacobi formalism, and then used for a consistent canonical
quantization, with the correct classical limit.\end{quote}

\bigskip\noindent{\large{\bf 1 \ Introduction}}\bigskip

Classical field theories describe physical phenomena by means of field
variables subject to partial differential equations. It often happens
that the number of field variables exceeds that of the physical degrees
of freedom: there is no unambiguous way of prescribing the values of the
field variables that correspond to a given physical situation. In the
mathematical structure of the theory, this property is reflected by the
existence of a gauge group, that allows transformations of the field
variables while the physical situation remains unchanged.

The peculiar feature of Einstein's theory of gravitation, which sets it
quite apart from ordinary field theories, is that its gauge group
consists of arbitrary distortions of the space-time coordinates, and
thus cannot be disentangled from the structure of space-time itself. In
particular, the time evolution of the gravitational field is locally
indistinguishable from a gauge transformation---namely a local
distortion of the space-time coordinates. As a consequence, the
Hamiltonian density $\cal H$, which generates the time evolution of the
field variables, is {\it weakly\/} equal to zero [1]: namely, although
$\cal H$ is a nontrivial function of the field variables (so that it can
generate a nontrivial evolution), its {\it numerical\/} value is
constrained to vanish.

This Hamiltonian constraint does not cause any difficulty in the
classical canonical theory. The numerical value of the Hamiltonian is
only an initial value constraint. It is the {\it functional\/} form
of the Hamiltonian that is needed for deriving the equations of motion.
In quantum theory, however, if the gravitational field equations are
quantized according to the standard canonical rule, namely
$\pi^{mn}=-i\hbar\delta/\delta g_{mn}$, the resulting Wheeler-DeWitt
equation [2,~3] leads to a dilemma known as ``the problem of time.'' The
difficulty is that, when the constraint $H\Ps=0$ is imposed on the state
vector $\Ps$, the latter is ``frozen.'' There cannot be wavepackets
moving along classical trajectories, in accordance with Ehrenfest's
theorem [4]. That is, {\it the quantum equations do not lead to the
expected classical limit\/}.

The problem was investigated long ago by Arnowitt, Deser, and Misner~[5]
who showed that there is an infinite number of possible coordinate
conditions that may be used to put the theory in canonical form. The
imposition of these coordinate conditions is equivalent to the
introduction of ``intrinsic'' coordinates, defined by the dynamical
variables of the physical system. The ADM method~[5] and similar
ones~[6] provide, in principle, a completely general solution to the
problem. Unfortunately, it is difficult to actually implement such a
solution for a given, specific, physical situation.

In particular, many authors have been interested in the properties of
highly sym\-metric cosmological models, for which there is a reasonable
hope of obtaining an explicit solution. The trouble is that these
symmetric situations have fewer (if any) dynamical degrees of freedom to
which the ADM conditions can be applied, and the problem of time arises
again. It would be impossible to mention here all the attempts that were
made to solve that problem. Only a few randomly chosen references are
listed below [7--11], with apologies to the authors of many similar
works.

Why is general relativity special? The reason is the totalitarian nature
of Einstein's theory: {\it all\/} physical phenomena are coupled to
gravitation, for the simple reason that all phenomena occur in
space-time, and the properties of space-time are determined by the
gravitational field. Any stresses, or any other physical forces, are
themselves sources of the gravitational field, by virtue of the Einstein
equations. (Electro\-magnetic theory, as a counter\-example, is
compatible with the existence of forces of non-electro\-magnetic nature,
that have no electro\-magnetic field of their own, and can cause
electric charges to move in arbitrary ways.)

On the other hand, quantum theory, unlike general relativity, is {\it
not\/} a ``theory of everything'' [12]. Its mathematical formalism can
be given a consistent physical interpretation only by arbitrarily
dividing the physical world into two parts: the system under study,
represented by vectors and operators in a Hilbert space, and the
observer (and the rest of the world), for which a classical description
is used. This point was emphasized long ago by Bohr~[13]:

\begin {quote}The necessity of discriminating in each experimental
arrangement between those parts of the physical system which are to be
treated as measuring instruments and those which constitute the objects
under investigation may be said to form a {\it principal distinction
between classical and quantum-mechanical description of physical
phenomena\/} \ldots\ The place within each measuring procedure where
this discrimination is made is largely a matter of
convenience.\end{quote}

\noindent The consistency of this hybrid quantum-classical formalism can
formally be proved, under suitably restrictive assumptions on the
properties of the classical world [14]. That is, as foreseen by Bohr,
the precise location of the boundary between the classical and quantum
parts of the system is irrelevant for well posed problems.

General relativity and quantum theory therefore appear to be
fundamentally incompatible. Nevertheless, as will be shown in this
article, they can be combined in a consistent way, by a careful choice
of the dynamical variables. This is shown below by means of two simple
examples, with a few degrees of freedom. Each example starts by
specifying a Lagrangian.  Indeed it is known that canonical commutation
relations are compatible with specified equations of motion only if the
latter are equivalent to the Euler-Lagrange equations derived from some
Lagrangian [15]. In both examples, the Lagrangians are chosen in such a
way that the resulting dynamics are afflicted by the infamous ``problem
of time,'' just as in canonical quantum gravity. Yet, canonical
quantization is possible, provided that one degree of freedom is kept
classical, so that it can be used as a clock.

The general method for solving this type of problem, for an arbitrary
Lagrangian, is briefly discussed. It involves the solution of a first
order partial differential equation, similar to the Hamilton-Jacobi
equation, but with a very different physical meaning. \\[6mm]

\noindent{\large{\bf 2 \ A simple example of constrained 
dynamics}}\bigskip

As our first example, consider a dynamical system with three degrees of
freedom, $x, y, z$, and with a Lagrangian

\beq
 L={1\over2}\left({\dot{x}^2\over z}-zx^2-{\dot{y}^2\over z}+zy^2\right).
\eeq
The canonical momenta are $p_x=\dot{x}/z$, \ $p_y=-\dot{y}/z$, and
$p_z=0$. The equations of motion thus are

\beq \dot{p}_x=d(\dot{x}/z)/dt=\partial L/\partial x=-zx, \label{px} 
\eeq
\beq \dot{p}_y=d(-\dot{y}/z)/dt=\partial L/\partial y=zy, \label{py}
\eeq
\beq \dot{p}_z=0=\partial L/\partial z=\half\,(-p_x^2-x^2+p_y^2+y^2).
\label{pzdot} \eeq
Equations of motion with a similar behavior occur in a cosmological
model with a homogeneous scalar field, as in the Weinberg-Salam theory,
but without a Higgs potential~[16].

It is now convenient to introduce an auxiliary time, $\tau$, by means of
$d\tau=zdt$, so that $p_x=dx/d\tau$ and $p_y=-dy/d\tau$. Together with
this new time, we also have a new Lagrangian, given by $L_\tau
d\tau=Ldt$, so that the action remains the same. That is, 

\beq L_\tau=L/z={1\over2}\left[\left({dx\over d\tau}\right)^2-x^2 
  -\left({dy\over d\tau}\right)^2+y^2\right]. \eeq
The corresponding Hamiltonian is

\beq H_\tau=\half\,(p_x^2+x^2-p_y^2-y^2), \label{Ham} \eeq
and it is easy to derive from it the equations of motion of the two
harmonic oscillators, $x$ and $y$. It follows from Eq.~(\ref{pzdot})
that $H_\tau=0$. This causes no difficulty at the classical level. The
Hamilton equations of motion are derived from the functional form of the
Hamiltonian, irrespective of its numerical value, and they are
equivalent to Eqs.~(\ref{px}) and (\ref{py}) above.

Trouble arises, however, if we attempt to quantize such a system by
introducing a wave function $\psi(x,y)$ that satisfies $H_\tau\psi$=0.
Separation of variables readily leads to the general solution

\beq \psi=\sum_n c_n\,u_n(x)\,u_n(y), \eeq
where the $c_n$ are arbitrary constants, and the $u_n$ are harmonic
oscillator eigenfunctions corresponding to energy $E_n=
(n+{1\over2})\hbar$. Obviously, this state is time independent. Nothing
moves. If we try to get a semi-classical solution by using large values
of $n$, we find that the amplitude of the wave function is large in the
vicinity of the four corners of a square, $x,\,y\simeq\pm\sqrt{2E_n}$,
and it is of course fixed in time.

On the other hand, there is no such difficulty with the Heisenberg
equations of motion, for example, $dx/d\tau=[x,H_\tau]/i\hbar$. These are
formally identical to the classical oscillator equations of motion, and
they lead to a nontrivial motion of the Heisenberg operators. We thus
see that the Ehrenfest theorem~[17] and, more generally, the
correspondence principle, are not valid for such a dynamical system.

In order to find a quantum counterpart to the dynamical system that is
represented, in classical physics, by the Hamiltonian (\ref{Ham}), we
must proceed more carefully. One of the harmonic oscillators, for
example $y$, will serve us as a clock, and then the other one can be
quantized in the usual way. We thus perform, still at the classical
stage, a canonical trans\-formation from $y$ and $p_y$ to new canonical
variables,

\beq Q^0=\tan^{-1}(p_y/y), \eeq

\beq P_0=-(p_y^2+y^2)/2. \eeq
There is no corresponding unitary transformation in quantum mechanics
(since the spectrum is not invariant), but in classical mechanics, such
a canonical trans\-formation is perfectly possible.

It is easily seen that $[Q^0,H_\tau]\pb=1$, so that

\beq dQ^0/d\tau=1. \label{Qeqtau} \eeq
We can thus write 

\beq H_\tau=P_0+H=0, \label{Htau} \eeq
where $H=\half\,(p_x^2+x^2)$ is the ordinary Hamiltonian of the
$x$-oscillator. Its equations of motion are $dx/d\tau=p_x$ and
$dp_x/d\tau=-x$. Thanks to Eq.~(\ref{Qeqtau}), we can also write them
as

\beq dx/dQ^0=p_x, \eeq

\beq dp_x/dQ^0=-x. \eeq
Finally, if we replace $p_x$ by $-i\hbar\partial/\partial x$, and $P_0$
by $-i\hbar\partial/\partial Q^0$, as usual, Eq.~(\ref{Htau}) becomes
the standard Schr\"odinger equation for a harmonic oscillator.

However, at this point, we must be careful: the wave function
$\psi(x,Q^0)$ should be normalized according to

\beq \int|\psi(x,Q^0)|^2\,dx=1, \eeq
without any further integration $\int\cdots\,dQ^0$. This follows from
our decision of keeping the clock time $Q^0$ classical, so that it can
play the ordinary role of time in Schr\"odinger's equation. In this way,
we have obtained a simple, consistent formalism, with the correct
classical limit.  Obviously, there are many other possible consistent
formalisms, that are not equivalent to each other, and yet give the same
classical limit.  Quantization is possible, but it is not a unique
process.\\[6mm]

\noindent{\large{\bf 3 \ Definition of a dynamical time}}\bigskip

It will be now be shown how a similar quantization process can be
performed for any classical dynamical system with a constrained
Hamiltonian,

\beq H_\tau(q,p)=0. \eeq
Here, $q$ and $p$, without indices, mean $\{q^1\ldots q^n\}$ and
$\{p_1\ldots p_n\}$, respectively, and $\tau$ is an arbitrary,
convenient time parameter, in terms of which the problem has been
formulated. In the case of Einstein's gravitational field equations,
there is an infinite number of dynamical variables and of constraints
(there are twelve canonical variables, $g_{mn}$ and $\pi^{mn}$, and four
constraints per space point). In the present article, however, my main
interest is in solving the ``problem of time'' for mini\-super\-spaces.
I shall therefore assume that there is only a finite number, $n$, of
degrees of freedom, and a single Hamiltonian constraint,
Eq.~(\theequation).

Following the method illustrated in the preceding section, let us seek a
canonical trans\-formation from $\{q^k\}$ and $\{p_k\}$ to new canonical
variables, $\{Q^\mu\}$ and $\{P_\mu\}$, with $\mu=0,\ldots,n-1$, such
that

\beq dQ^0/d\tau=[Q^0,H_\tau]\pb=1. \eeq
It follows from Eq.~(\theequation)  that

\beq H_\tau=P_0+H(Q^0\ldots Q^{n-1}, P_1\ldots P_{n-1}).\label{Hcan}\eeq
The latter equation defines an effective Hamiltonian $H$. Note that $H$
does not depend on $P_0$ so that, after we replace $P_0$ by
$-i\hbar\partial/\partial Q^0$, the classical equation $H_\tau=0$
becomes a Schr\"odinger equation for the new time $Q^0$ and the $(n-1)$
dynamical variables, $Q^1\ldots Q^{n-1}$.

The first step thus is to find a suitable clock time $Q^0(q,p)$.  This
can easily be done, as least in a restricted domain of phase space, as
shown in Fig.~1. We start with an arbitray $(2n-1)$-dimensional
hyper\-surface $\cal K$, oriented in such a way that the flow lines
$dq/d\tau=\partial H_\tau/\partial p$ and $dp/d\tau=-\partial
H_\tau/\partial q$ are nowhere tangent to $\cal K$. That is, all the
flow lines lie on the same side of $\cal K$, for $\tau$ positive and
short enough. Then, at least for some finite time, these flow lines will
not intersect---as long as they do not reach a critical point---and they
will not reenter $\cal K$ from the other side (however, if the motion is
bounded, for example if it is periodic, reentry must obviously happen
after enough time has elapsed). Anyway, for a finite time, each flow
line that originates from $\cal K$ ascribes a unique set of $q$ and $p$
to each value of $\tau$ and, conversely, in a finite domain of phase
space, there is a unique $\tau$ for each set of $\{q, p\}$, say
$\tau=f(q, p)$.

There still is here a formidable technical difficulty, because generic
dynamical problems are not integrable: the number of constants of motion
is usually less than the number of degrees of freedom, and the function
$f(q,p)$ defined above cannot be obtained in closed form and does not
exist globally. For that reason, some authors [18] take the liberty of
``fixing the gauge'' by an arbitrary choice of the function $f(q,p)$,
leading to a form which is convenient for further work. It is not clear
to me why this is permitted. This is also not necessary, because there
do exist approximation methods for performing a sequence of canonical
transformations which reduce the Hamiltonian to a normal form~[19,~20].
These give approximate constants of motion, which are represented in
phase space by ``vague tori'' and are useful for describing the dynamics
over extended time periods. These tori remnants~[21,~22] become
important in quantum theory because, if their missing parts are small
compared to $2\pi\hbar$, the quantum system behaves as if it were
regular, with ordinary selection rules.

\begin{figure}[t]
\includegraphics{wdw.fig} 
\vspace{82mm}\hspace*{23mm}\parbox{113mm}{FIG. 1. \ Orbits in phase
space, starting on the hypersurface $\cal K$. Each one of these orbits
defines an internal clock-time, $Q^0(q,p)$.} \end{figure}

Let us now return to the ``problem of time.'' What we need is a
canonical transformation such that $Q^0=f(q, p)$ is a prescribed
function. As explained above, $f(q,p)$ is chosen in such a way that

\beq [f(q,p), H_\tau]\pb=1, \eeq
and therefore $dQ^0/d\tau=1$. At this point, it is natural to ask what
would happen if we had chosen another initial hypersurface, say $\cal
K'$, leading us to a different time function, $f'(q,p)$, say. We would
then have

\beq [f(q,p)-f'(q,p), H_\tau]\pb=0, \eeq
so that $(f-f')$ has to be a constant of the motion. Either it is a
function of $H_\tau$ or, if there are other, non\-trivial constants of
the motion, $(f-f')$ can be a function of them. In particular, if
$H_\tau=0$ is the only constant of motion, $(f-f')$ can only be a mere
number. Anyway, it does not matter for the sequel whether whether $f(q,
p)$ is uniquely defined, up to a numerical constant, or can be modified
by adding a non\-trivial constant of the motion (thus effectively giving
a different version of the theory).

Our task thus is is to find explicitly a canonical transformation that
leads to the decomposition~(\ref{Hcan}). This can be done, in principle,
by the solution of a first order partial differential equation of the
same type as the Hamilton-Jacobi equation. It is easiest to use a
generating function~[23] of type $F_1$, that we shall write as $S(q,
Q)$. We have

\beq p_k=\partial S/\partial q^k, \label{p} \eeq

\beq P_\mu=-\partial S/\partial Q^\mu. \label{P}\eeq
Since $S$ is time-independent (there is no explicit appearance of
$\tau$ in $S$), the new Hamiltonian is numerically equal to the old one,
as in Eq.~(\ref{Hcan}).

To obtain $S$ explicitly, we have to solve

\beq Q^0=f\left(q,{\partial S(q,Q)\over\partial q}\right).\label{HJ}\eeq
The various $Q^\mu$, with $\mu>0$, are unspecified integration constants
in the solution of Eq.~(\theequation). As in the Hamilton-Jacobi case,
there is no guarantee that (\theequation) has well behaved global
solutions. However, it is always possible to achieve arbitrarily close
approximations in a finite domain. An example is given in the next
section.

Once we have obtained $S(q,Q)$, we get $P_\mu(q,Q)$ from Eq.~(\ref{P}).
We can then invert these equations, in principle, and find $q(Q,P)$.
Likewise, Eq.~(\ref{p}) gives us $p=p(q,Q)$, and since $q(Q,P)$ is
already known, this gives $p=p(Q,P)$. All these results are then
substituted in $H_\tau$ so as to obtain the explicit form of
Eq.~(\ref{Hcan}). Finally, that equation can be quantized in the usual
way, replacing $Q^0$ by a new variable, $t$ (recall that
$dQ^0/d\tau=1$), and $P_0$ by $-i\hbar\partial/\partial t$. Note,
however, that the new parameter $t$ is not a function of the space-time
coordinates:  it is a function, $f(q,p)$, of the phase-space
coordinates. This is a meaningful dynamical time, not a meaningless
(gauge dependent) coordinate-time.\\[6mm]

\noindent{\large{\bf 4 \ Quantization of a minisuperspace}}\bigskip

Let us finally return to general relativity. As the simplest example,
consider a spatially flat Friedmann-Lema\^\i tre universe~[24], with
metric

\beq ds^2=N^2(t)\,dt^2-a^2(t)\,(dx^2+dy^2+dz^2). \eeq
The matter source is a massless scalar field $\phi$, for which the
energy density and pressure are

\beq \rho=p=\half\,\dot{\phi}^2, \eeq
where natural units have been used: $c=8\pi G=1$.

The Einstein field equations, for the above metric and sources, become
ordinary differential equations for the three variables $N(t)$, $a(t)$,
and $\phi(t)$. In order to obtain a quantum version of this theory, the
above differential equations must be obtainable as the Euler-Lagrange
equations resulting from a Lagrangian~[15]. It is easily found that a
suitable Lagrangian, giving the correct equations, is

\beq L=(\half\,a^3\,\dot{\phi}^2-3a\,\dot{a}^2)/N. \eeq
Note that $\dot{N}$ does not appear in $L$, so that

\beq p_N\equiv\partial L/\partial\dot{N}=0, \eeq
and therefore

\beq \dot{p}_N=\partial L/\partial N=-L/N^2=0.\label{Leq0}\eeq
The fact that $L=0$ is an initial value constraint imposed on the
variables $a$, $\dot{a}$, and $\dot{\phi}$.

Likewise, we have

\beq p_\phi\equiv\partial L/\partial\dot{\phi}=a^3\,\dot{\phi}/N. \eeq
This is a constant of the motion, because $\partial L/\partial\phi=0$.
Finally,

\beq p_a\equiv\partial L/\partial\dot{a}=-6\,a\,\dot{a}/N, \eeq
and

\beq \dot{p}_a=\partial L/\partial a=
  3\,(\half\,a^2\,\dot{\phi}^2-\dot{a}^2)/N. \eeq

As in Sect.~2, it is convenient to introduce an auxiliary time $\tau$ by
means of $d\tau=Ndt$. We then have a new Lagrangian, given by $L_\tau
d\tau=Ldt$, so that the action remains invariant. Furthermore, it is
convenient to introduce, instead of the radial scale variable $a$, a new
variable, $v(t)=a^3(t)$, which scales the volume element. We then have

\beq L_\tau={v\over2}\,\left({d\phi\over d\tau}\right)^2-
 {1\over3v}\left({dv\over d\tau}\right)^2, \label{Ltau}\eeq
from which we obtain

\beq p_\phi=v\,{d\phi\over d\tau}, \eeq

\beq p_v=-{2\over3v}\,{dv\over d\tau}. \label{pv}\eeq
The corresponding Hamiltonian is

\beq H_\tau\equiv p_\phi\,{d\phi\over d\tau}+p_v\,{dv\over d\tau}-L_\tau
 ={1\over2v}\,p_\phi{}^2-{3v\over4}\,p_v{}^2.\label{H4}\eeq
Note that both $L_\tau$ and $H_\tau$ vanish weakly, as a consequence of
(\ref{Leq0}). The non-essential dynamical variable $N(t)$ has thus been
eliminated, but it has left a remnant, which is the initial value
constraint, $H_\tau=0$. The equations of motion resulting from the new
Hamiltonian are: $p_\phi=$ const., Eq.~(\ref{pv}), and

\beq {dp_v\over d\tau}=-{\partial H_\tau\over\partial v}=
 {p_\phi{}^2\over2v^2}+{3\,p_v{}^2\over4}. \label{pvt} \eeq

Our task now is to find a dynamical time function,
$Q^0=f(v,p_v,p_\phi)$, such that $dQ^0/d\tau=1$. (Obviously, $f$ is not
a function of the cyclic variable $\phi$, since the latter does not
appear explicitly in the equations of motion.) In other words, we want
a function $f(v,p_v,p_\phi)$ that satisfies

\beq [f(v,p_v,p_\phi), H_\tau]\pb =1. \label{fH}\eeq
For this, we have to solve the equations of motion explicitly.

Substitution of (\ref{pv}) into the right hand side of (\ref{H4}) gives,
after some rearrangement,

\beq \half\,(dv/d\tau)^2+\2\,H_\tau v=\4\,p_\phi{}^2. \label{vt}\eeq
This looks like the elementary energy equation for free fall of a
particle of unit mass, height $v$, and total energy $\4\,p_\phi{}^2$, in
a gravity field $g=\2H_\tau$. The solution is

\beq v=-\4\,H_\tau\,\tau^2\pm\sqrt{\2}\,p_\phi\,\tau,\label{v}\eeq
where the integration constant was set so that $v=0$ when $\tau=0$
(in other words, the $\cal K$ hyper\-surface is given by $v=0$). Since
by definition $v\geq0$, the $\pm$ sign in (\theequation) has to be the
same as the sign of of $p_\phi\tau$. Note that we are not allowed to set
$H_\tau=0$ at this stage: consistency of the method that was proposed in
the preceding section requires that the equations of motion be valid for
the entire phase space, not only for the orbits with initial conditions
that satisfy $H_\tau=0$.

It is possible to solve directly (\theequation) for $\tau$, and then to
substitute (\ref{H4}) in the result. However, it is simpler to proceed
as follows. From (\theequation), we have

\beq {dv\over d\tau}=-\2\,H_\tau\,\tau\pm\sqrt{\2}\;p_\phi,\eeq
whence, thanks to Eq.~(\ref{pv}),

\beq v\,p_v=-\3\,{dv\over d\tau}=H_\tau\,\tau\mp\sqrt{\3}\;p_\phi. \eeq
Thus, (\ref{v}) becomes

\beq v=-\4\,\tau\,(H_\tau\,\tau\mp\sqrt{\7}\,p_\phi)
 =-\4\,\tau\,(v\,p_v\mp\sqrt{\3}\,p_\phi),\label{vpv}\eeq
and therefore

\beq \tau\equiv f(v,p_v,p_\phi)=
  {v\over -\4\,v\,p_v\pm\sqrt{\8}\,p_\phi}\,.\label{tau}\eeq
It is easy to verify directly that Eq.~(\ref{fH}) indeed holds.

The next step is to find explicitly the transformation from the original
canonical coordinates to the new ones, that include $Q^0$ and $P_0$. We
have, from Eqs.~(\ref{HJ}) and (\ref{tau}),

\beq Q^0={v\over-\4\,v\,(\partial S/\partial v)\pm
  \sqrt{\8}\,(\partial S/\partial\phi)}, \label{S}\eeq
where $S=S(v,\phi,Q^0,Q^1)$. An obvious way for obtaining a solution is
to separate variables, namely,

\beq S=\phi\,Q^1+S'(v,Q^0,Q^1), \eeq
so that

\beq Q^1=p_\phi.\eeq
Rearranging Eq.~(\ref{S}), we obtain

\beq {1\over Q^0}=-\4\,{\partial S'\over\partial v}\pm
 \sqrt{\8}\,{Q^1\over v}, \eeq
whose solution is

\beq S'= \9\,\left(-{v\over Q^0}\pm\sqrt{\8}\; Q^1\,\ln v\right). \eeq
We thus have

\beq p_v={\partial S'\over\partial v}=\4\,\left(-{1\over Q^0}
 \pm\sqrt{\8}\,{p_\phi\over v}\right), \eeq
in agreement with (\ref{vpv}).

Note that

\beq P_0=-\partial S'/\partial Q^0=-\9\,v/(Q^0)^2,\eeq
whence

\beq v=-\4\,P_0\,(Q^0)^2. \eeq
When these equations for $v$ and $p_v$ are substituted into (\ref{H4}),
we obtain

\beq H_\tau=P_0\pm\sqrt{\7}\,p_\phi/Q^0. \eeq

The reduced Hamiltonian $H$, defined by Eq.~(\ref{Hcan}), thus is

\beq H=\pm\sqrt{\7}\,p_\phi/Q^0.\label{HQ}\eeq
Recall that the $\pm$ sign in $H$ is the same as the sign of
$p_\phi/Q^0$. Note that if $Q^0$ is considered as equivalent to the time
$\tau$, the number of degrees of freedom has been reduced by~2: the
variable $N$ disappeared in the trans\-formation from $t$ to $\tau$, and
the $v$ and $p_v$ variables have been absorbed in the dynamical
definition of a ``clock-time'' $Q^0$.

Here, we must be careful and avoid expressing $Q^0$, in
Eq.~(\theequation), by means of the right hand side of (\ref{tau}). This
would give

\beq H=(p_\phi{}^2/v)\mp\sqrt{\2}\,p_\phi\,p_v\qquad{\rm(wrong)}.
 \label{Hv}\eeq
Such a way of writing the Hamiltonian is not correct: it would give the
true equations of motion for $v$ and $p_v$ only if the initial
conditions are set in such a way that $H_\tau=0$ in Eq.~(\ref{H4}),
namely

\beq p_\phi{}^2=\2\,(vp_v)^2. \label{constr}\eeq
Indeed, we have from (\ref{Hv})

\beq dv/d\tau=[v,H]\pb=\mp\sqrt{\2}\,p_\phi\qquad{\rm(wrong)}, \eeq
and

\beq dp_v/d\tau=[p_v,H]\pb=p_\phi{}^2/v^2\qquad{\rm(wrong)}, \eeq
and these agree with Eqs.~(\ref{pvt}) and (\ref{vt}) only if
(\ref{constr}) is satisfied. Therefore $H$ in Eq.~(\ref{Hv}) is not a
valid, unconstrained Hamiltonian for this problem. Only $H$ given by
(\ref{HQ}) is acceptable. (More generally, the reader may easily verify
that $P_0=H_\tau-H$ has vanishing Poisson brackets with all the
canonical variables only on the hypersurface $H_\tau=0$.)

We thus remain with the reduced Hamiltonian (\ref{HQ}), and we may now
replace in it $Q^0$ by $\tau$. The only nontrivial equation of motion is

\beq {d\phi\over d\tau}=[\phi,H]\pb=\pm\sqrt{\7}\,{1\over\tau}, \eeq
whence $\phi=\phi_0\pm\sqrt{\7}\ln\tau$. Quantization is trivial: the
wave function $\psi(p_\phi,\tau)$ satisfies a Schr\"odinger equation,

\beq i\hbar\,{\partial\psi\over\partial\tau}=
  \pm\sqrt{\7}\,{p_\phi\over\tau}\,\psi, \eeq
so that

\beq \psi=
 F(p_\phi)\,\exp\left(\mp{i\over\hbar}\sqrt{\7}p_\phi\ln\tau\right),\eeq
where $F(p_\phi)$ is an arbitrary function that takes care of
normalization. 

Obviously, only a superspace with a larger number of degrees of freedom
can give an interesting theory. Unfortunately, ``interesting'' also
means ``non\-integrable'': the function $f(q,p)$ is not in general well
behaved (it is not ``isolating'') and approximation methods must be
used~[19--22].

Finally, the question must be raised whether the notion of a
mini\-super\-space is a valid approximation for studying quantum
gravity~[25]. The arbitrary imposition of symmetry constraints on the
gravitational field freezes almost all its dynamical degrees of freedom
in a way that appears to be incompatible with the existence of quantum
fluctuations. A similar dilemma arises in elementary classical
mechanics, when we impose mundane mechanical constraints, such as
restricting the motion of a mass to a two-dimensional surface.
Classically, such a system is well defined. However, its quantization is
not unique and it essentially depends on the nature of the constraining
forces~[26]. I hope to return to this problem in a future publication.

\bigskip\noindent{\large{\bf Acknowledgment}}

\bigskip It is a pleasure to dedicate this article to Englebert
Sch\"ucking, on the occasion of his 70th birthday. I am grateful to
Stanley Deser for clarifying comments. This work was
supported by the Gerard Swope Fund, and the Fund for Encouragement of
Research.%\clearpage

\bigskip\noindent{\large{\bf References}}

\frenchspacing
\begin{enumerate}
\item P. A. M. Dirac, Proc. Roy. Soc. (London) A 246 (1958) 333.
\item J. A. Wheeler, in {\it Battelle Rencontres: 1967 Lectures on
Mathematical Physics\/} (Benjamin, New York, 1968).
\item B. S. DeWitt, Phys. Rev. 160 (1967) 1113.
\item T. Brotz and C. Kiefer, Nucl. Phys. B 475 (1996) 339.
\item R. Arnowitt, S. Deser, and C. W. Misner, in {\it Gravitation: an
Introduction to Current Research\/}, ed. by L.~Witten (Wiley, New York,
1962).
\item A. Peres, Phys. Rev. 171 (1968) 1335.
\item W. G. Unruh, Phys. Rev. D 40 (1989) 1048.
\item W. G. Unruh and R. M. Wald, Phys. Rev. D 40 (1989) 2598.
\item C. G. Torre, Phys. Rev. D 46 (1992) 3231.
\item R. M. Wald, Phys. Rev. D 48 (1993).
\item J. D. Brown and K. Kucha\v r, Phys. Rev. D 51 (1995) 5600. 
\item A. Peres and W. H. Zurek, Am. J. Phys. 50 (1982) 807.
\item N. Bohr, Phys. Rev. 48 (1935) 696.
\item A. Peres, {\it Quantum Theory: Concepts and Methods\/} (Kluwer,
Dordrecht, 1993) p.~376.
\item S. A. Hojman and L. C. Shepley, J. Math. Phys. 32 (1991) 142.
\item V. N. Pervushin and V. I. Smirichinski, report JINR E2-97-155
(e-print gr-qc/9704078).
\item P. Ehrenfest, Z. Phys. 45 (1927) 455.
\item M. Cavagli\`a, V. de Alfaro, and A. T. Filippov, Int. J. Mod.
Phys. A10 (1995) 611.
\item G. Contopoulos, Astrophys. J. 138 (1963) 1297.
\item F. G. Gustavson, Astronom. J. 71 (1966) 670.
\item C. Jaff\'e and W. P. Reinhardt, J. Chem. Phys. 77 (1982) 5191.
\item R. B. Shirts and W. P. Reinhardt, J. Chem. Phys. 77 (1982) 5204.
\item H. Goldstein, {\it Classical Mechanics\/} (Addison-Wesley,
Reading, MA, 1980) p.~382.
\item C. W. Misner, K. S. Thorne, and J. A. Wheeler, {\it Gravitation\/}
(Freeman, San Francisco, 1973) Chapt.~27.
\item K. V. Kucha\v r and M. P. Ryan, Jr., Phys. Rev. D 40 (1989) 3982.
\item N. G. van Kampen and J. J. Lodder, Am. J. Phys. 52 (1984) 419.

\end{enumerate}\end{document}